\documentclass[aps,prl,twocolumn,showpacs,superscriptaddress,groupedaddress]{revtex4-1}  % for review and submission
\usepackage{graphicx}  % needed for figures
\usepackage{dcolumn}   % needed for some tables
\usepackage{bm}        % for math
\usepackage{amssymb}   % for math

\begin{document}

% The following information is for internal review, please remove them for submission
%\widetext
%%\leftline{Primary authors: Joe E. Physics}
%\leftline{To be submitted to PRX}
%\leftline{Comment to {\tt d0-run2eb-nnn@fnal.gov} by xxx, yyy}
%\centerline{\em D\O\ INTERNAL DOCUMENT -- NOT FOR PUBLIC DISTRIBUTION}

% the following line is for submission, including submission to the arXiv!!
%\hspace{5.2in} \mbox{Fermilab-Pub-04/xxx-E}

\title{Spatial coherence and stability in a disordered organic polariton condensate}

\author{K. S. Daskalakis}
\author{S. A. Maier}%

\affiliation{Department of Physics, Imperial College London, London, SW7 2AZ, United Kingdom}

\author{S. K\'{e}na-Cohen}%
\affiliation{Department of Engineering Physics, \'{E}cole Polytechnique de Montr\'{e}al, Montr\'{e}al, QC, H3C 3A7, Canada}
\email{s.kena-cohen@polymtl.ca}

%\input author_list.tex       % D0 authors (remove the first 3 lines
                             % of this file prior to submission, they
                             % contain a time stamp for the authorlist)
                             % (includes institutions and visitors)
\date{\today}

\begin{abstract}
Although only a handful of organic materials have shown polariton condensation, their study is rapidly becoming more accessible. The spontaneous appearance of long-range spatial coherence is often recognized as a defining feature of such condensates. In this work, we study the emergence of spatial coherence in an organic microcavity and demonstrate a number of unique features stemming from the peculiarities of this material set. Despite its disordered nature, we find that correlations extend over the entire spot size and we measure $g^{(1)}(r,r')$ values of nearly unity at short distances and of 50\% for points separated by nearly 10~$\mu$m. We show that for large spots, strong shot to shot fluctuations emerge as varying phase gradients and defects, including the spontaneous formation of vortices. These are consistent with the presence of modulation instabilities. Furthermore, we find that measurements with flat-top spots are significantly influenced by disorder and can, in some cases, lead to the formation of mutually incoherent localized condensates.
\end{abstract}

\maketitle

%\section{\label{sec:level1}First-level heading}
% sections are not used for PRL papers

In the last few years, a number of molecular systems that show room-temperature polariton condensation have emerged \cite{Kena-Cohen2010,Daskalakis2014,Plumhof2014}. Polaritons are hybrid exciton-photon quasiparticles that form in optical microcavities when dissipation is low as compared to the the light-matter interaction. Like their fundamental constituents, they obey Bose statistics at low densities. They inherit an effective mass as low as $10^{-10}$ times that of a Rb$^{87}$ atom from their photonic component. Meanwhile, they collide with other polaritons and excitons due to an effective interaction stemming from their matter component. In inorganic microcavities, these features have been exploited to demonstrate rich phenomenology associated with Bose-Einstein condensation (BEC). The most well-known of these effects is the macroscopic occupation of the ground state beyond a critical density $n_c$. A perhaps more important consequence of the phase transition is the sudden appearance of off-diagonal long range order (ODLRO) in coordinate space \cite{Penrose1956,Yang1962}. This manifests itself in the non-vanishing long-range behaviour of the first-order spatial coherence $g^{(1)}(\textbf{r},\textbf{r'})$. At the polariton condensation threshold, a similar transition occurs, thus breaking U(1) symmetry, but also showing a number of distinct features resulting from the strongly non-equilibrium nature of the system~\cite{Wouters2007}. 

Several materials have been used to demonstrate polariton condensation. The majority of these have used CdTe- and GaAs-based semiconductors whose operation is limited to low temperatures due to their small exciton binding energy. Recently, however, wide bandgap semiconductors such as ZnO and GaN have emerged as viable materials for room-temperature applications \cite{Lu2012,Li2013,Daskalakis2013,Christopoulos2007}. Organic semiconductors are especially attractive in this context due to their large exciton binding energy. Molecular excitons are commonly of the Frenkel type where both electron and hole are localized on a single molecule and exciton transport occurs incoherently between neighbouring molecules. In microcavities, long-range coupling between molecules is provided by the photon field and theoretical studies have shown that a phase transition can occur even in the absence of electronic coupling between molecular dipoles \cite{Keeling2004}. Organics offer the advantage of a broad spectral range beyond that covered by GaN and ZnO and can be easily fabricated without the need for epitaxial growth. As-grown films, however, tend to be highly disordered and to overcome possible localization effects ~\cite{Michetti2005,Marchetti2006,Malpuech2007,Malpuech2012,
Zajac2012,Savona2007,Houdre2000,Borri2000,Gurioli2001,Janot2013,Thunert2014,Baas2008,Agranovich2003}, the first demonstration of organic polariton condensation used a single-crystalline organic semiconductor as the active material~\cite{Kena-Cohen2010}.
 Two recent demonstrations, however, have shown that the same phenomenology can be extended to amorphous systems: one consisting of a spin-coated polymer \cite{Plumhof2014} and the other of a thermally evaporated oligomer \cite{Daskalakis2014}. By using interferograms, these reports have also evidenced the presence of long-range correlations in these systems. Here, we study in detail the onset of spatial and temporal coherence in an organic polariton condensate. We show that even in the presence of disorder, correlations can span the entire system size. Moreover, we highlight the important role of the pump and observe a number of unique features such as the spontaneous formation of vortices, which may suggest the presence of modulation instabilities.

\begin{figure}
\includegraphics[scale=0.3]{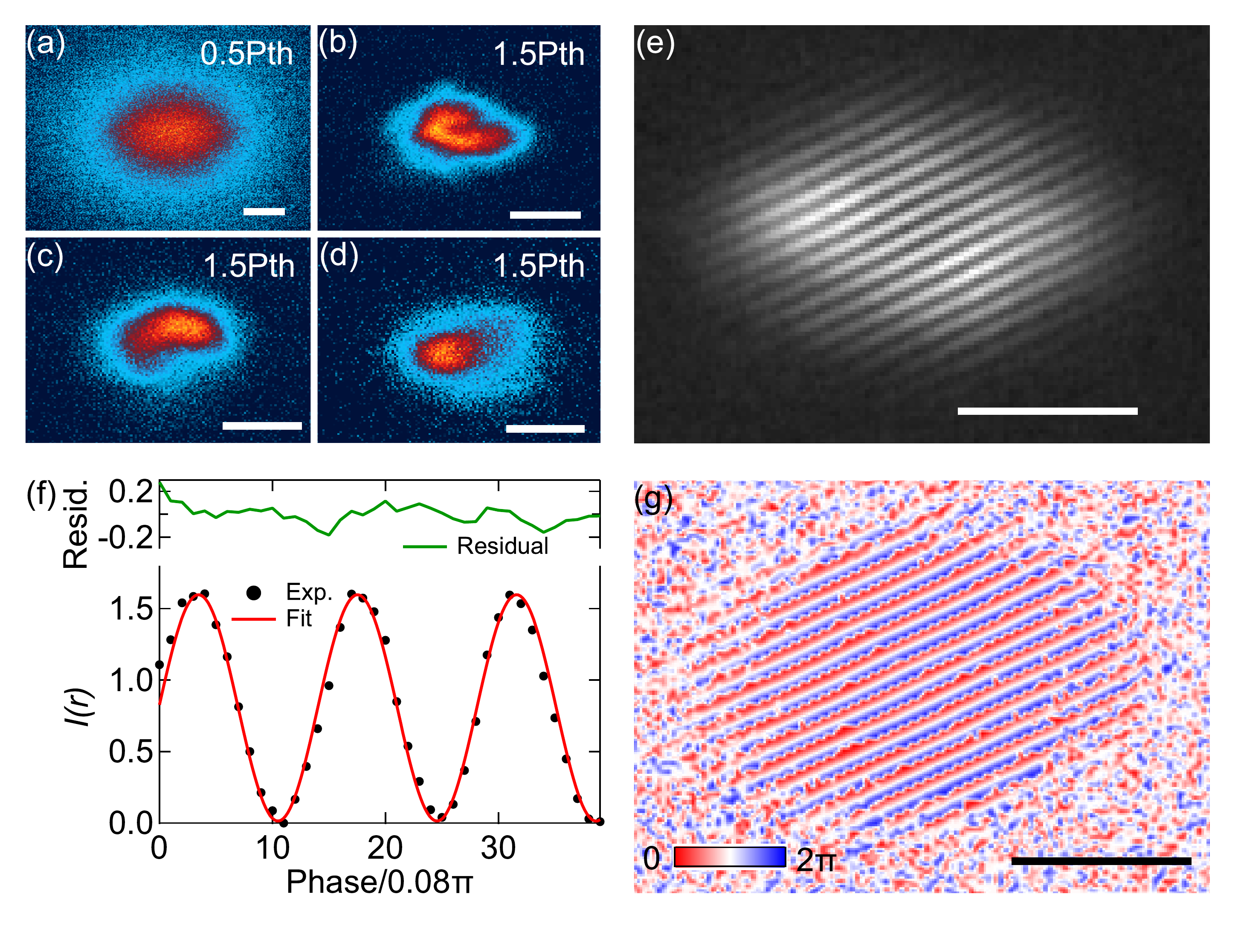}
\caption{\label{fig:epsart} (color online). (a-d) Real-space images of the time-integrated PL below (a) and above threshold at different locations (b--d) using an elliptical Gaussian pump. The linear image (a) remains identical regardless of the sample location. (e) Interferogram from location (d) averaged over 15 laser shots from. (f) Typical intensity versus phase delay. For every pixel, a cosine fit is used to extract both $g^{(1)}(-\textbf{r},\textbf{r})$ and the phase offset. (g) phase extracted from the fit. Scale bar is 5~$\mu$m.}
\end{figure}

The microcavity is identical to that of Ref.~\cite{Daskalakis2014}. It uses 9 dielectric mirror pairs on opposite sides of a layer of 2,7-bis[9,9-di(4-methylphenyl)-fluoren-2-yl]-9,9-di(4-methylphenyl)fluorene (TDAF) and was impulsively pumped high above the polariton energy (i.e. nonresonantly). To highlight the presence of disorder, real-space photoluminescence (PL) using a Gaussian pump is shown in Fig.~\ref{fig:epsart}. In the linear regime, given in Fig.~\ref{fig:epsart}(a), we find that the PL is independent of sample location. Above threshold, however, in-plane potential energy fluctuations cause strong variations as a function of sample location, as shown in Fig.~\ref{fig:epsart}(b--d). Although elastic scattering due to disorder has been previously observed for organic polaritons in the linear regime~\cite{Kena-Cohen2008}, the local intensity fluctuations seen here only appear beyond the condensation threshold. These bear striking similarities to those observed in disordered inorganic microcavities~\cite{Krizhanovskii2009,Baas2008,Kasprzak2006,Zajac2012}.

Although a number of other criteria exist, the spontaneous onset of long-range spatial coherence is often recognized as a defining feature for condensation. This was first reported for a CdTe microcavity in the seminal work of \citeauthor{Kasprzak2006} using a Michelson interferometer with one arm replaced by a retroreflector~\cite{Kasprzak2006}. Soon after, the spatial coherence of a GaAs microcavity was measured using a Young's double slit experiment and found to reach 30\% at distances of 8 $\mu$m \cite{Deng2007}. Theoretical predictions regarding the asymptotic behaviour of $g^{(1)}(\textbf{r},\textbf{r'})$ have shown that it can be used to shed light on the nature of the phase transition \cite{Szymanska2006} and experiments with a low noise floor have observed a power-law decay and attributed this to a Berezinskii-Kosterlitz-Thouless (BKT) phase transition~\cite{Roumpos2012}. Recent theoretical work has, however, suggested that this decay may be the result of a crossover phenomenon at intermediate length scales and that correlations should instead fall-off exponentially~\cite{Altman2015}. Interferometric techniques have also been used to identify excitations such as vortices and vortex-antivortex pairs and to probe quantum fluidic effects in polaritonic systems \cite{Roumpos2010,Lagoudakis2011,Lagoudakis2008}.

\begin{figure*}
\includegraphics[scale=0.35]{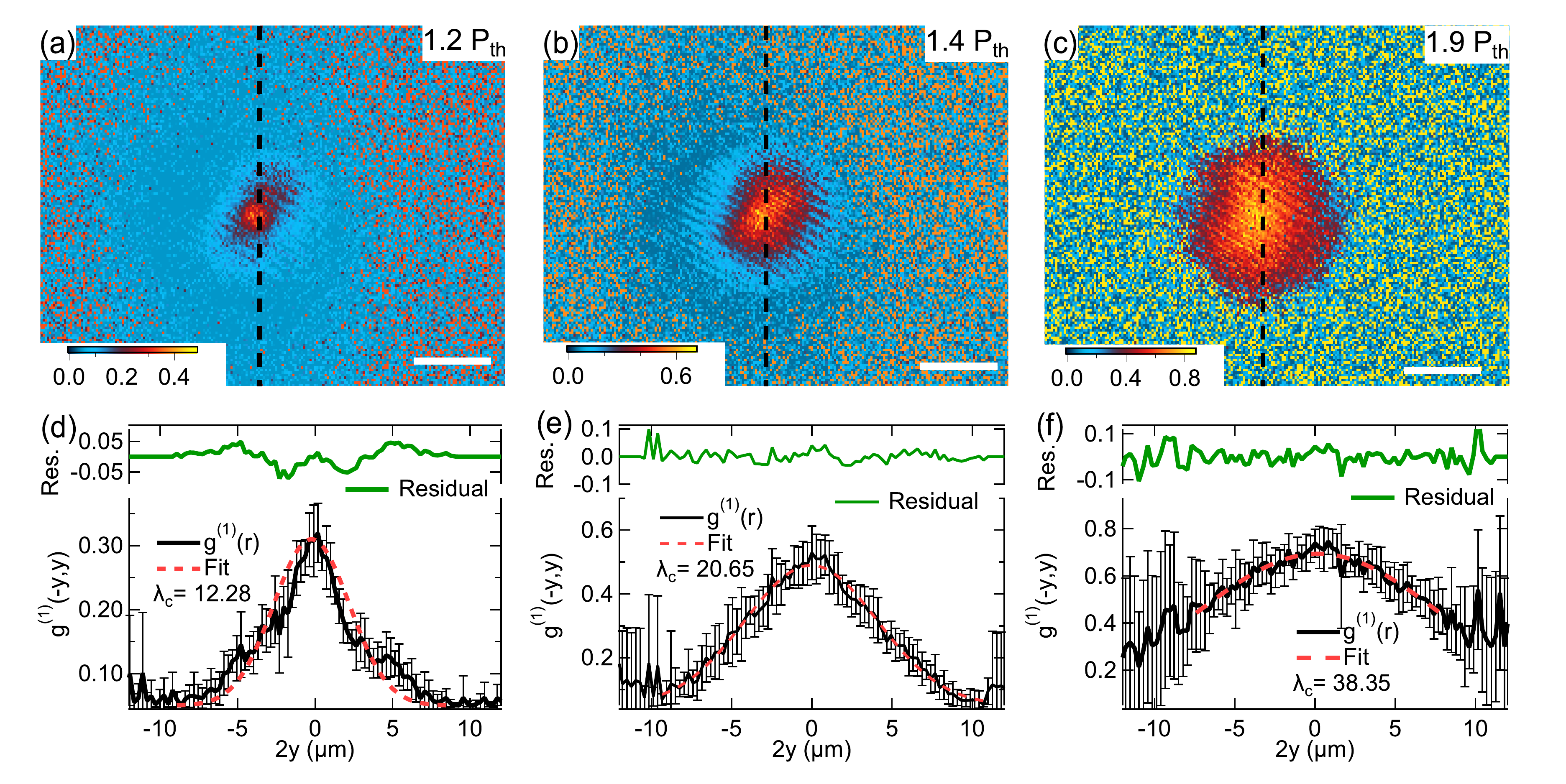}
\caption{\label{fig:epsart1} (color online). (a -- c) Contour maps of $g^{(1)}(-\textbf{r},\textbf{r})$ for increasing pump fluence. The colorbar maximum corresponds to the peak value of $g^{(1)}(\textbf{r},-\textbf{r})$ in the panel. (d -- f) Vertical cuts through the (a -- c) colour maps used to extract $g^{(1)}(-y,y)$. These are averaged over 8 horizontal pixels, with the standard deviation shown as error bars. The resulting coherence length is labelled $\lambda_c$ and the fit residual is shown. Scale bars are 5~$\mu$m.}
\end{figure*}

In this work, we use the retroreflector configuration from Ref.~\cite{Kasprzak2006}. A typical interferogram measured at the location of Fig.~\ref{fig:epsart}(d) is shown in Fig.~\ref{fig:epsart}(e). Despite the non-uniform intensity, parallel fringes indicate a flat phase over the entire condensate area. Similar results are observed for all sample locations. The visibility and phase offset are extracted by scanning a delay mirror (DM) over a phase of $6\pi$ and fitting the resulting pattern at each pixel as shown in Fig.~\ref{fig:epsart}(f). The offset is shown in Fig.~\ref{fig:epsart}(g) and is due to the small angle difference between the two interferometer arms. 
For a circular spot, one expects ballistic propagation of polaritons away from the center accompanied by a radially increasing local wavevector for the condensate~\cite{Wouters2008}. An upper limit for this propagation length can be calculated using $l_p\lesssim \hbar k_c / \gamma_c  m^* = 2$~$\mu$m, where $k_c=$~2.1~$\mu$m$^{-1}$ is the largest condensate wavevector supported by the dispersion, $m^*=2.1 \times 10^{-5}m_e$ is the effective mass and $\gamma_c=6$~ps$^{-1}$ is the polariton decay rate. For small enough pump spots, this leads to an annular condensate density as shown in the Supplementary Material (SM).

\begin{figure}
\includegraphics[scale=0.37]{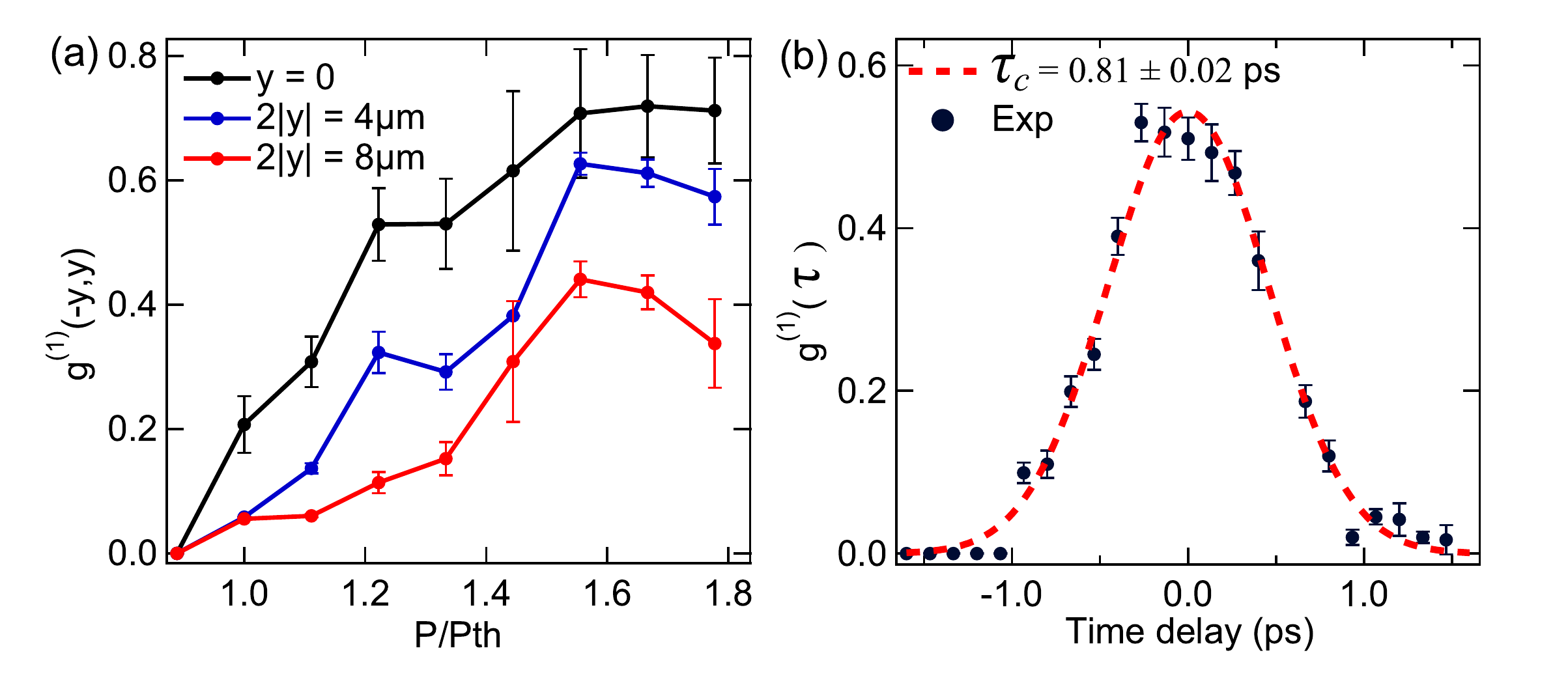}
\caption{\label{fig:epsart2} (a) The spatial coherence for increasing pump fluence. We find a sharp increase in $g^{(1)}(r,-r)$, followed by a plateau near $P=1.5P_{th}$. Additional details are provided in the SM. (b) First-order temporal coherence, $g^{(1)}(\tau)$, for $P=1.7P_{th}$, measured for two points separated by $3~\mu$m. Values of $g^{(1)}(\tau)$ are shown as solid black circles and the error bars indicate the standard deviation for each point. The temporal coherence of the condensate, $\tau_{c}$, obtained from a Gaussian fit (red dashed line) is indicated as $\tau_{c}$.}
\end{figure}

Figure~\ref{fig:epsart1} shows the spatial dependence of $g^{(1)}(-\textbf{r},\textbf{r})$ for 3 values of pump fluence. The spatial coherence reaches peak values of $g^{(1)}(-\textbf{r},\textbf{r})>0.8$ for points separated by up to 5~$\mu$m. Vertical cuts, averaged over 8 pixels through the center of Fig.~\ref{fig:epsart1}(a -- c) are shown in Fig.~\ref{fig:epsart1}(d -- f). A vertical profile was chosen to avoid artefacts resulting from the small collapse along the horizontal direction. These occur due to the small difference in the arrival time of the $50^{\circ}$ pump pulse along this axis. The noise remains significant near the condensate edge, which prevents a clear statement regarding the functional dependence of the asymptotic behaviour of $g^{(1)}(-\textbf{r},\textbf{r})$. In analogy with the thermal case, we simply show a Gaussian fit, which shows a reasonable agreement and allows us to extract a coherence length $\lambda_c=2\sqrt{2\pi}\sigma$, where $\sigma$ is the standard deviation of the Gaussian fit. We obtain a coherence length of up to $\lambda_c=38~\mu$m, which is ultimately limited by the pump size.  A typical power dependence of $g^{(1)}(-y,y)$ is shown in Fig.~\ref{fig:epsart2}(a). In all cases, an increase in $g^{(1)}(-y,y)$ is observed, followed by a plateau near $1.5P_{th}$. 

As previously reported, the effective polariton-polariton and polariton-exciton nonlinearities in state-of-the-art organic microcavities are dominated by exciton saturation. The resulting change in refractive index causes anti-guiding in analogy with gain guided, index anti-guided conventional lasers. Moreover, the resulting nonlinearities are much weaker than in inorganics, which are dominated by the Coulomb exchange term between carriers~\cite{Ciuti1998}.  From the blueshift in the region where the reservoir occupation is clamped (see SM), we can extract a polariton-polariton interaction coefficient $g = 10^{-6}$ meV $\mu$m$^2$. This allows us to estimate a typical healing length, calculated at $1.5P_{th}$, $\xi=\hbar/ \sqrt{m^* g |\Psi|^2}=2~\mu$m, where $|\Psi|^2$ is the condensate density. This sets the scale for density fluctuations and is in reasonable agreement with our observations. Although the effect of disorder is apparent in the intensity patterns, the long coherence length and high degree of first-order coherence suggests the presence of a single condensate with a large condensate fraction \cite{Wouters2008}. 

For an equilibrium condensate, increasing density typically leads to a screening of the disorder potential. \citeauthor{Thunert2014} \cite{Thunert2014} have recently reported on the important role of dissipation in non-equlibrium polariton condensates. They show that the competition between gain and dissipation can prevent screening of the disorder expected when the polariton interaction dominates over the potential fluctuations. Similarly in this work, we do not observe a screening of potential fluctuations up to a blueshift of $\mu\sim$~8~meV.
 
\begin{figure}
\includegraphics[scale=0.5]{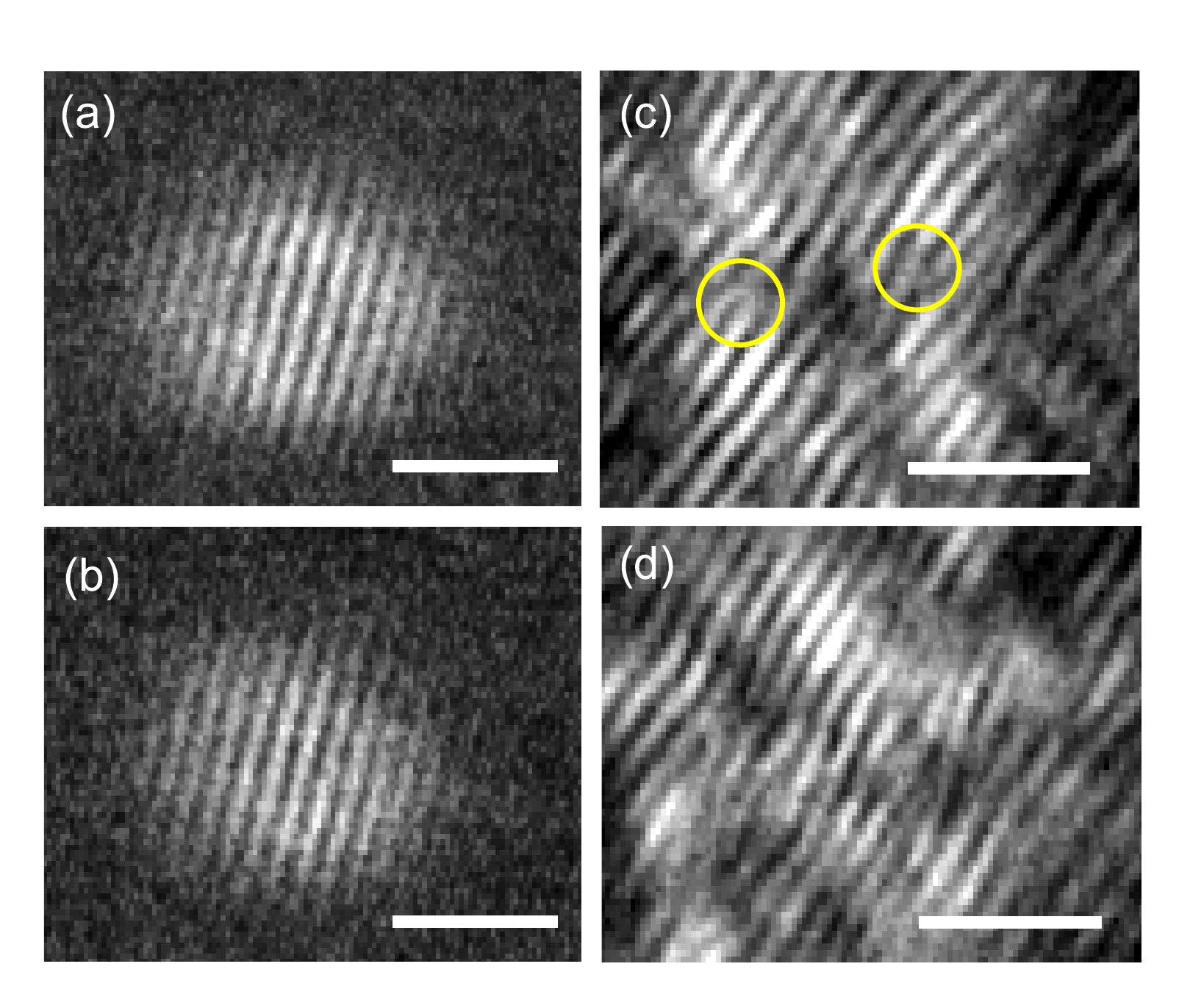}
\caption{\label{fig:epsart3} (color online). (a -- b) Interferograms using a small (27 x 25~$\mu$m) Gaussian pump taken at the same sample location and power integrated over 4 laser shots. (c--d) Interferograms using a large (97 x 60 $\mu$m) Gaussian pump integrated over 2 laser shots. Yellow circles indicate a fork dislocation. Scale bar is 5~$\mu$m.}
\end{figure}   

The same interferometry setup was used to measure the first-order temporal coherence of the polariton condensate. Here, the decrease in fringe contrast is measured over long time delays for two points spatially separated by $3~{\mu}$m. The temporal coherence, shown in Fig.~\ref{fig:epsart2}(b), exhibits a Gaussian profile with a coherence time $\tau_{c} = 0.8~$ps. Due to the impulsive pump, this value is not dominated by fluctuations, but instead agrees well with our calculation of the condensate survival time. From the emission linewidth, we find a condensate decay time of $1~$ps, which is in good agreement with the measured temporal coherence.

\begin{figure}
\includegraphics[scale=0.4]{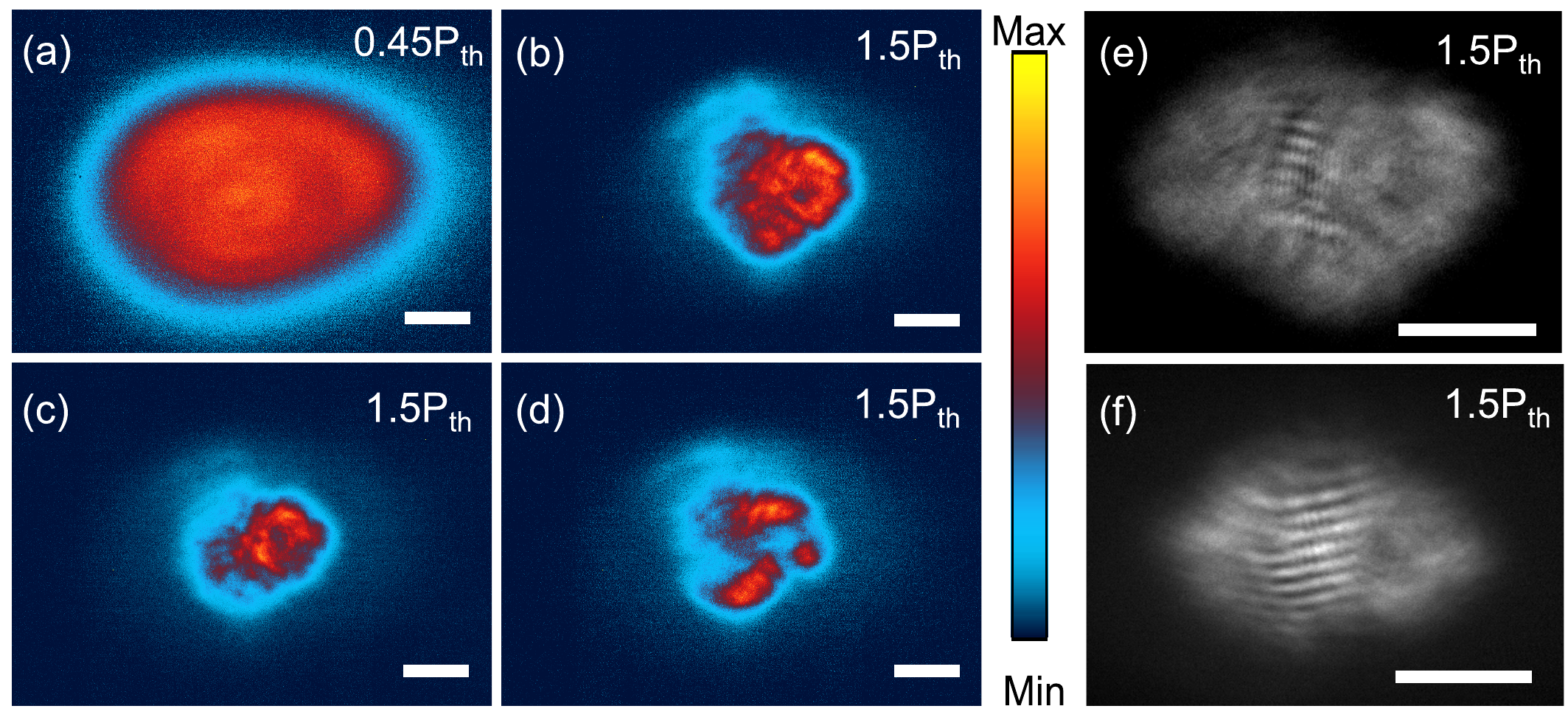}
\caption{\label{fig:epsart4} (color online). Real-space emission for a flat-top pump spot below (a) and above threshold (b -- d). (c) and (d) are intensity patterns that were recorded from different sample locations. (e -- f) interference patterns recorder at the same pump power (1.5P$_{th}$) and represent the sample locations (b) and (c), respectively. Note the change in fringe orientation, cross-fringes and poor contrast. Scale bar is 15~$\mu$m.}
\end{figure}

One of the most startling features occurs for larger spot sizes. We find that for powers greater than  $\sim 1.5P_{th}$ phase gradients (density currents) and dislocations are commonly observed. These change from shot-to-shot and the overall interference pattern becomes washed out. This is shown in Fig.~\ref{fig:epsart3} where few-shot images were taken at the same location with small and large Gaussian pumps at $1.8P_{th}$. Identical parallel fringes are observed for the small spot, but strong shot-to-shot variations occur for the larger pump. For a small fraction of the laser shots, centro-symmetric fork dislocations due to singly charged vortices such as those shown in Fig.~\ref{fig:epsart3}(c) can be observed. These form spontaneously at different locations within the spot from shot to shot. Real-space profiles for a large pump also show density fluctuations, while occasional regions of vanishing density can be attributed to vortex cores (see SM).

To emphasize the crucial role of the pump shape, we have performed similar experiments using a refractive beam shaper to control the pump beam profile. In this case, the excitation spot was made significantly larger and with a flat-top intensity. Figure~\ref{fig:epsart4} shows the below threshold (a) and above threshold (b -- d) real-space images from 3 different sample locations. Below threshold, the intensity is nearly flat over the entire pumped region, albeit with some small fluctuations. Above threshold, the appearance of multiple bright regions is observed. In some cases, these regions are separated by as much a ten microns. The resulting interferograms tend to be highly distorted often showing a mixture of dislocations, phase gradients and incoherent regions with strong shot-to-shot fluctuations. We have, in addition, verified that in cases with a significant separation between condensates, mutual coherence is lost (see SM).

Our experiments with various pump sizes and shapes point to two distinct causes for these distorted patterns. First, a greater sampling of the disorder for large spots can lead to condensation in regions further away from the center. In this case, our interferometric geometry which inverts about the origin will sample regions that can be uncorrelated or possess additional phase shifts due to their different positions. The latter is seen in Fig.~\ref{fig:epsart4}(e-f) where the fringe orientation changes despite the identical geometry. Note also the presence of smaller cross-fringes in both cases. These features are never seen for Gaussian pumping. Second, small Gaussian pumping has been theoretically predicted to be beneficial in reducing the region in reciprocal space where modulation instabilities can occur~\cite{Wouters2007,Smirnov2014,Bobrovska2014,Liew2015}. The criteria for such instabilities to occur is nearly always satisfied in organic cavities, which possess typical cavity decay rate $>10^2$ times those of reservoir excitons. In our case, for a homogeneous spot, these instabilities exist over a broad range of wavevectors for pump powers up to $~10^{4}P_{th}$. The short survival time of our condensate helps prevent their significant growth.

In conclusion, we find that organic polariton condensates bear similarities to disordered inorganic microcavities, but with some distinct features.  For smaller spots, we always observe a single condensate showing good phase correlation and a coherence length ultimately limited by the pump size. In contrast, for large Gaussian pumps, phase gradients and defects emerge, with strong shot-to-shot fluctuations. Although masked in the linear regime, the role of disorder becomes evident beyond the condensation threshold. For homogeneous spots, the interferograms become highly distorted as a result of \emph{both} disorder and fluctuations. The size and profile dependence are consistent with current predictions and suggest that the fluctuations may be due to modulation instabilities. Another interesting possibility, however, is that these may be remnants from a Kibble-Zurek-like mechanism~\cite{Liew2015}. The ease of fabrication of TDAF---it can be spin-coated or thermally evaporated---make it a useful platform for the further investigations of these effects.

\section{Acknowledgements}
The authors are grateful to Jonathan Keeling, Giuseppe La Rocca and Leonardo Mazza for providing insightful comments on the draft of this manuscript and Ray Murray for support. Work by K.D. and S.M. was supported by the Leverhulme Trust and EPSRC Active Plasmonics programme, while S.K.C. gratefully acknowledges support from the NSERC Disovery Grant progam.

\bibliography{MyCollection}% Produces the bibliography via BibTeX.
%\begin{thebibliography}{99}

%\end{thebibliography}

\newpage

\setcounter{equation}{0}
\setcounter{figure}{0}
\setcounter{table}{0}

\renewcommand{\theequation}{S\arabic{equation}}
\renewcommand{\thefigure}{S\arabic{figure}}

\onecolumngrid
\large
\section{\large{Supplementary Material}}

\subsection{\large{1. Sample and Experimental configuration}}

\begin{figure}[h]
\includegraphics[scale=0.55]{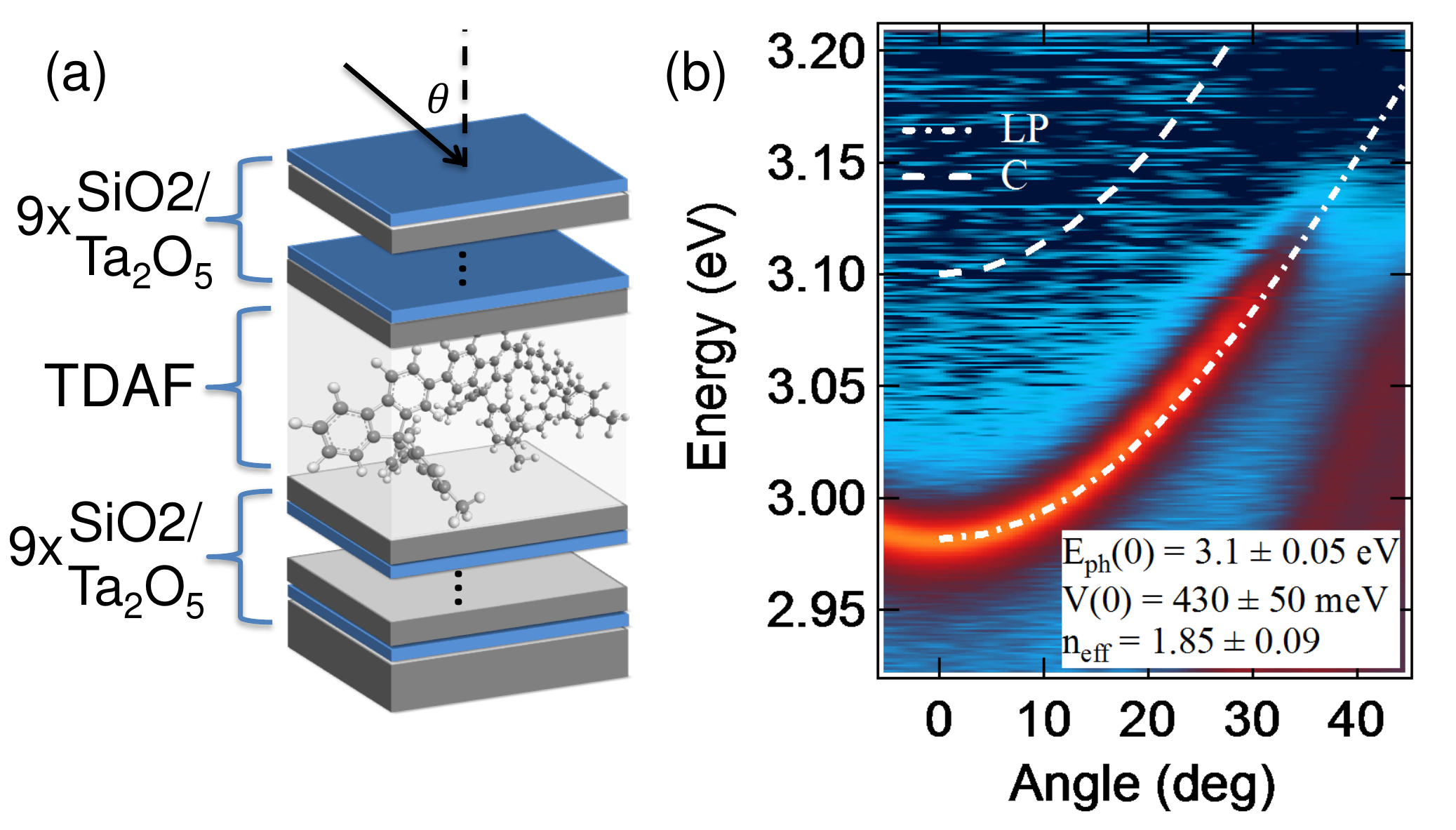}
\caption{\label{fig:supfig0} (color online). (a) Schematic of our microcavity. (b) TM-polarized PL of a negative detuned cavity ($\Delta$ = -400 meV). The detuning was obtained using the coupled harmonic oscillator model described in Ref.~\cite{Daskalakis2014}. Lower polariton (LP) and cavity mode (C) dispersions are shown with dashed lines and the fit parameters are indicated}
\end{figure}

As in our previous work, the microcavity used in this Letter consists of a 120 nm thick TDAF layer sandwiched between two 9-pair distributed Bragg reflectors (DBRs). These consist of alternating SiO$_{2}$ and Ta$_{2}$O$_{5}$ layers (Fig.~\ref{fig:supfig0}a). The sample is excited non-resonantly using 250~fs pulses from an optical parametric amplifier tuned to 3.22~eV. The pulses are TM-polarized and incident at 50$^{o}$ with a repetition rate of 1~kHz for PL and time-integrated interferometry and 100~Hz for the single-shot interferograms. Figure~\ref{fig:supfig0}b shows the TM-polarized PL measured on a location with a detuning $\Delta$ = -400~meV. The dispersion parameters are indicated in the figure.

\begin{figure}[h]
\includegraphics[scale=0.25]{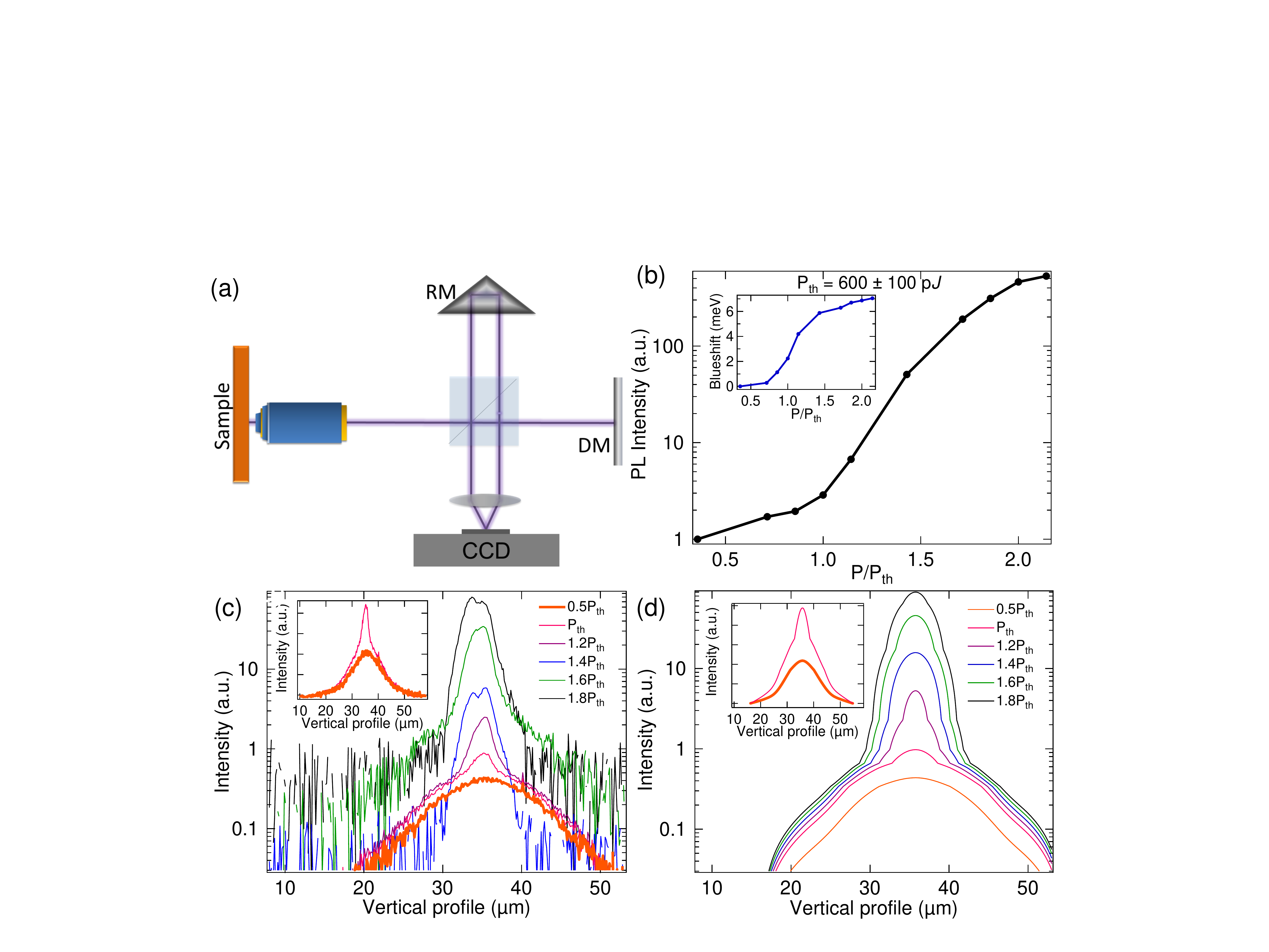}
\caption{\label{fig:supfig1} (color online). (a) Schematic of our experiment where one arm of a Michelson interferometer is replaced by a retroreflector (RM). A piezoelectric translation stage is to vary the phase introduced by the delay mirror (DM). The reflected images from RM and DM are then overlapped on a CCD camera (Thorlabs BC106N-VIS). (b) Power dependence of the PL and blueshift (inset) observed at normal incidence. (c) A typical vertical profile of the intensity for a small Gaussian pump as a function of increasing pump fluence. (d) Profile calculated by scaling the linear regime profile by the peak PL power dependence. The inset shows the profiles on a linear scale.}
\end{figure}

Our experimental configuration is shown schematically in Fig.~\ref{fig:supfig1}(a). In the retroreflector Michelson interferometer configuration, when the image centers are aligned, the first-order spatial coherence between centrosymmetric points $g^{(1)}(\textbf{r},\textbf{r'})=g^{(1)}(\textbf{r},-\textbf{r})$ can be obtained from the visibility, $V(r)$, of the interferogram fringes: 

\[\textit{V}(\textbf{r})= \frac{I_{max}-I_{min}}{I_{max}+I_{min}}= \frac{2\sqrt{I(\textbf{r})I(-\textbf{r})}}{I(\textbf{r})+I(-\textbf{r})}g^{(1)}(\textbf{r},-\textbf{r})\]

Where $I(\textbf{r})$ is the light intensity and $ I_{max} $ and $ I_{min} $ are respectively the maximum and minimum intensities measured, when the phase delay is varied. 

The power dependence and blue shift measured for our microcavity are given in Fig.~\ref{fig:supfig1}(b). The threshold energy is $P_{th}$ = 600~$\pm 100$~pJ, which corresponds to an absorbed pump fluence of $27\pm5$~$\mu$J/cm$^2$. The errors correspond to the standard deviation over various sample locations.

\subsection{\large{2. Large-scale collapse}}

Beyond threshold, we observe a collapse of the emission region towards the pump center ($\sim70\%$ reduction in full width at half maximum). A typical vertical intensity profile is shown in Fig.~\ref{fig:supfig1}(c) as a function of increasing pump power. We find that this large-scale collapse can be readily explained by scaling the intensity profile in the linear regime, which follows the pump profile, by the normal-incidence power dependence measured using a fiber-coupled CCD spectrometer shown in Fig.~\ref{fig:supfig1}(b). If $n(\textbf{r})$ is the normalized pump profile in the linear regime, $f(p)$ the power dependence at power $p=P/P_{th}$, the intensity is calculated such that $I(\textbf{r})=f\left(n\left(\textbf{r}\right)p\right)$. The agreement obtained is illustrated in Figure~\ref{fig:supfig1}(d). Note, however, that spatial disorder is \emph{not} captured by this model. The non-uniform density peaks within the condensate apparent in Fig.~\ref{fig:supfig1}(c) are absent in the linear profile and consequently in Fig. \ref{fig:supfig1}(d).

\subsection{\large{3. Outwards propagation of polaritons}}

\begin{figure}[h]
\includegraphics[scale=0.5]{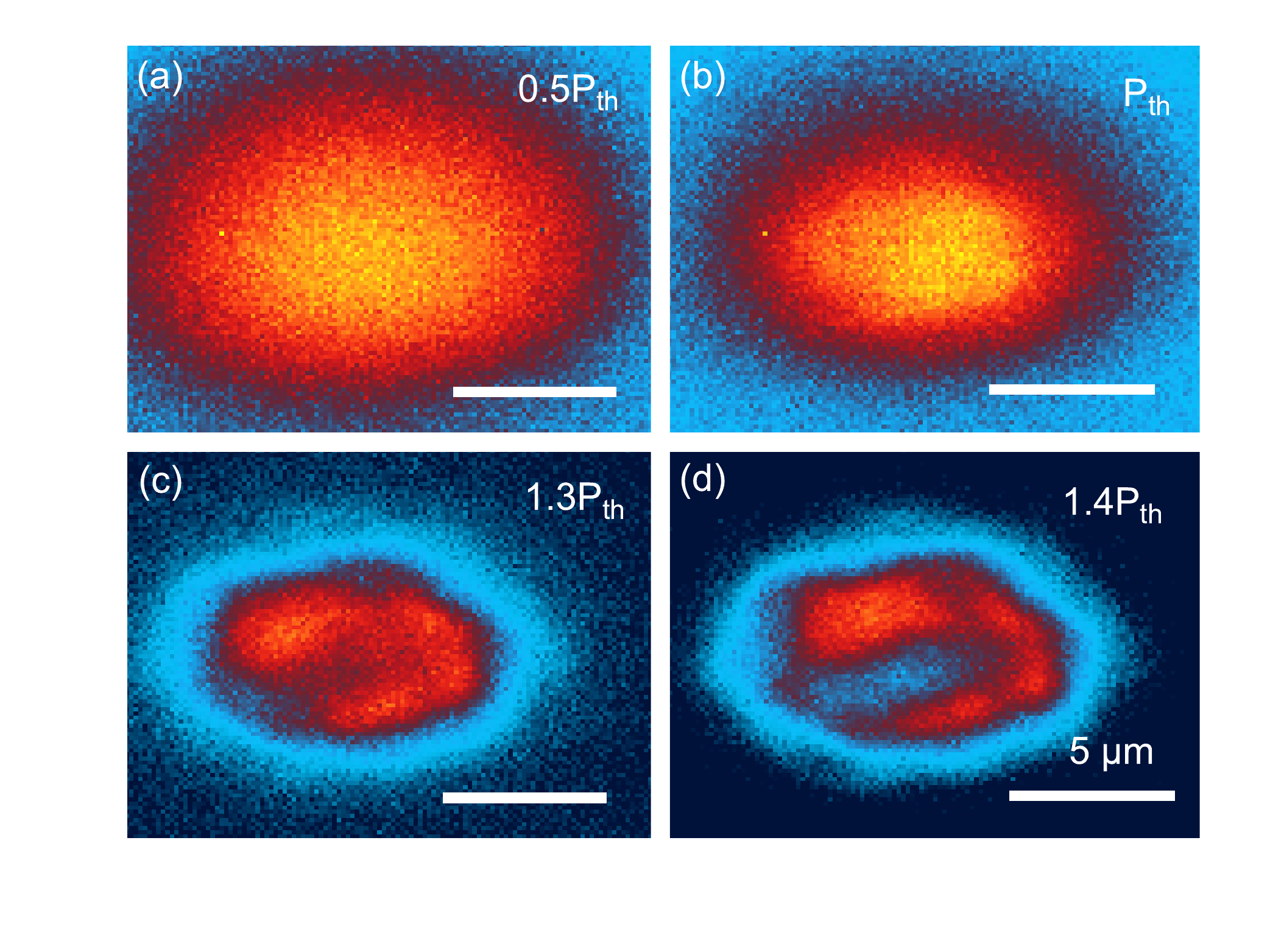}
\caption{\label{fig:repulsion} (color online). Typical real-space condensate emission measured as a function of increasing pump power for a small Gaussian pump}
\end{figure}

For small enough pump spots, the above threshold emission becomes distorted into a ring-like pattern in contrast to the homogeneous large-scale collapse. We attribute this to the outwards flow of polaritons due to the repulsive potential induced by the pump. This behaviour is shown in Fig.~\ref{fig:repulsion} and qualitatively observed everywhere on the sample. Disorder typically causes some deviation from a perfect ring-shape and we have carefully verified that this behaviour is reversible and not a result of damage within the central region.

\subsection{\large{4. Power-dependence of spatial coherence }}

\begin{figure}[h]
\includegraphics[scale=0.5]{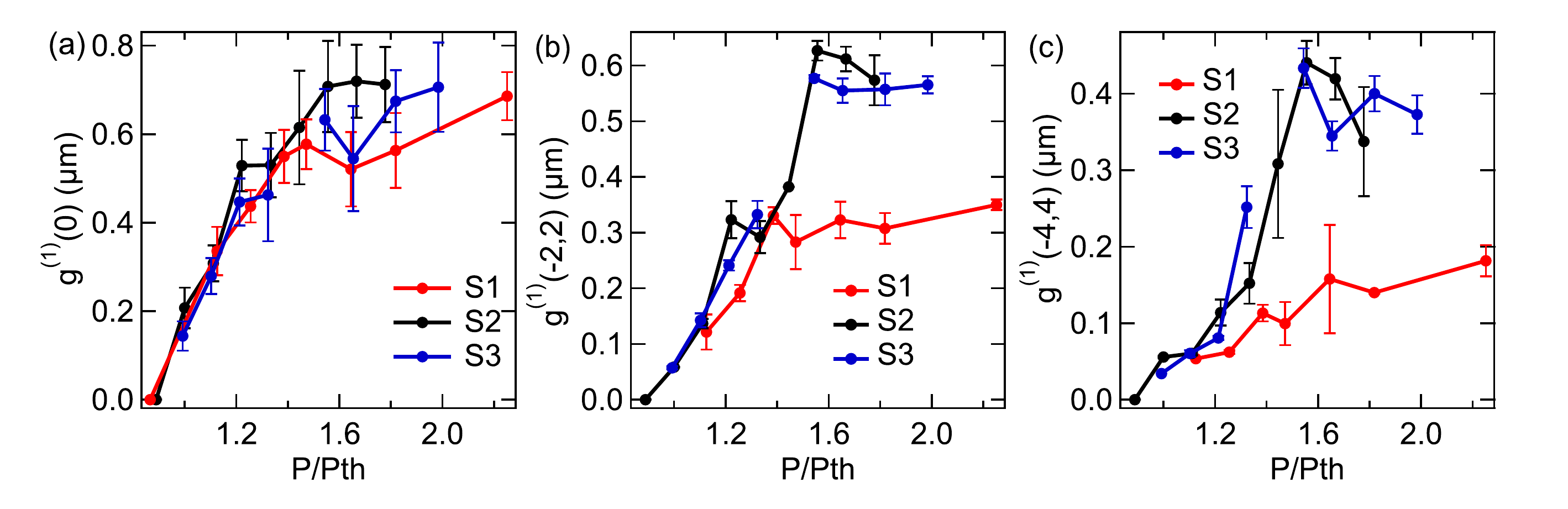}
\caption{\label{fig:supfig2} (color online). (a -- c) The spatial coherence at various locations away from the condensate center $r=0~\mu$m shown for increasing pump fluence. Note the sharp increase in $g^{(1)}(-y,y)$, followed by a plateau near $P=1.5P_{th}$, The value and error bars correspond to the mean and standard deviation obtained from averaging over 25 pixels.}
\end{figure}

We have measured the power-dependence of $g^{(1)}(-y,y)$ on three different sample locations using a small Gaussian pump. These are shown in Fig.~\ref{fig:supfig2} around $g^{(1)}(0,0)$ and at distances $|2y|=4~\mu$m and $|2y|=8~\mu$m. In all cases, values and standard deviations were obtained by averaging over 25 pixels on both sides of the origin.

\subsection{\large{5. Additional real-space and single-shot images}}

\begin{figure}[h]
\includegraphics[scale=0.4]{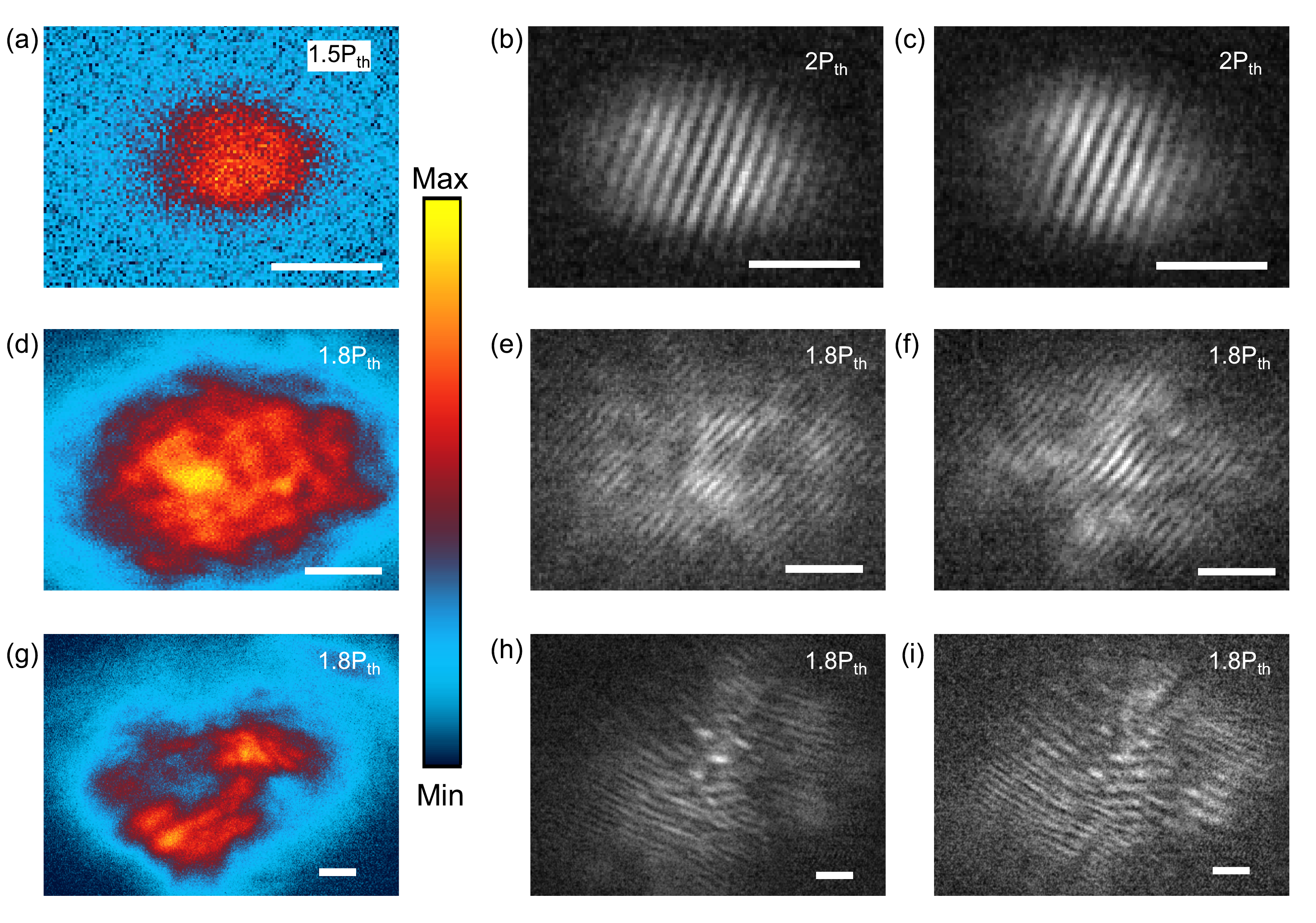}
\caption{\label{fig:supfig3} (color online). (a--c) Real-space (a) and 4-shot interferograms (b--c) for a small Gaussian pump. (d--f) Real-space (d) and 2-shot interferograms (e--f) for a large Gaussian pump. (g--i) Real-space (g) and 2-shot interferograms for a flat-top pump (h--i). Real space images were integrated over a) 100, d) 30 and g) 80 laser shots. Scale bar is 5~$\mu$m.}
\end{figure}

\begin{figure}[h]
\includegraphics[scale=0.5]{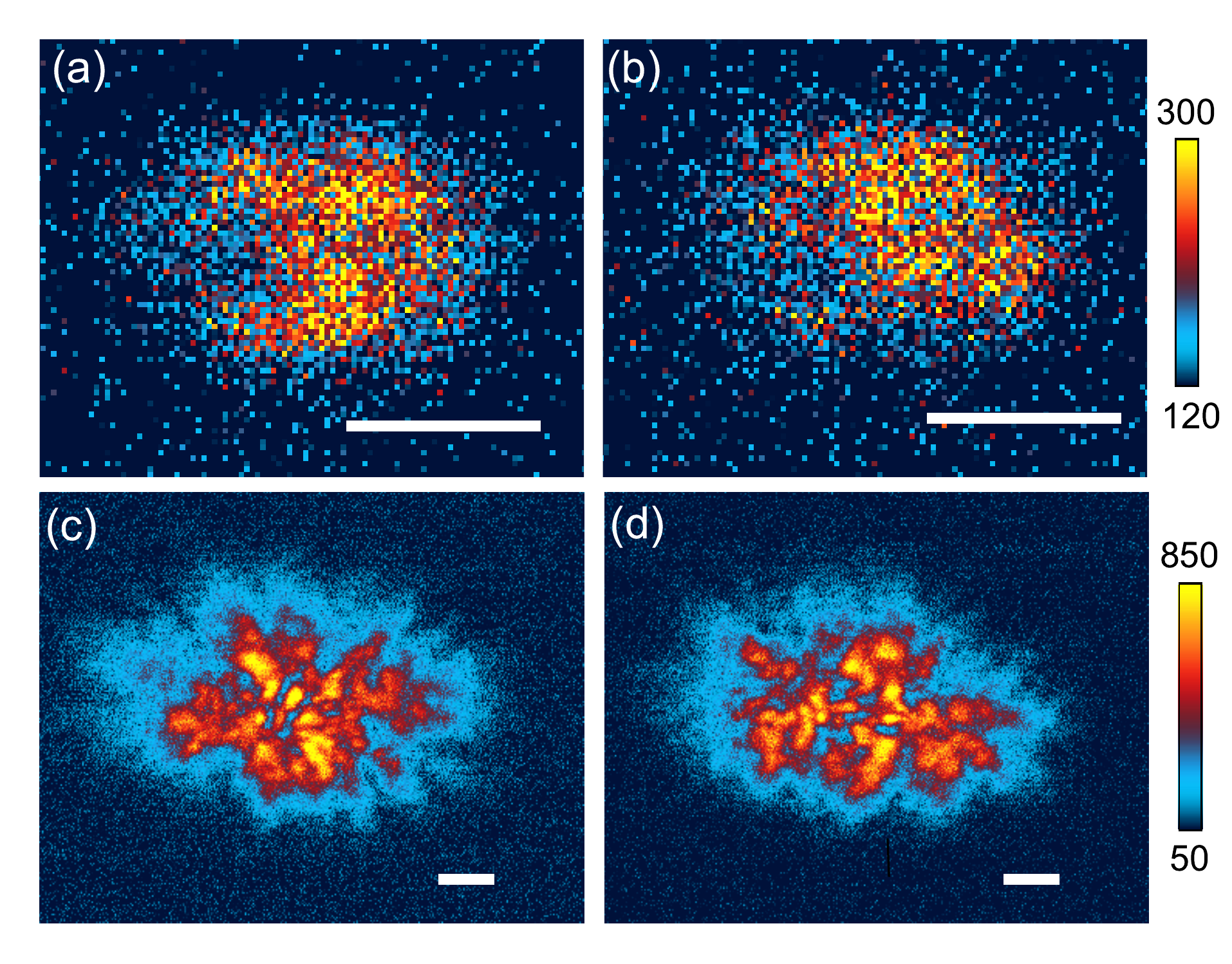}
\caption{\label{fig:supfig5} (color online). Two consecutive single-shot images taken above threshold for (a--b) a small Gaussian pump and (c--d) a large Gaussian pump. Scale bar is 5~$\mu$m.}
\end{figure}

Figure~\ref{fig:supfig3} shows supplementary few-shot interferograms images measured for the three different spot types. Averaged real-space images at the same location are included. Note the inhomogeneous character of the large Gaussian real-space image. We can also see that the flat-top pump results in a significantly larger number of phase defects as compared to a Gaussian pump. Single-shot real-space images comparing the behaviour for small and large pumps are shown in Figure~\ref{fig:supfig5}. Although we cannot simultaneously acquire a single-shot real-space image and interferogram in our setup, the occasional appearance of zero intensity regions in the real-space image with dimensions on the order of the healing length is consistent with the vanishing density of a vortex core. 

\subsection{\large{6. Mutual coherence between independent condensates}}

\begin{figure}[h]
\includegraphics[scale=0.7]{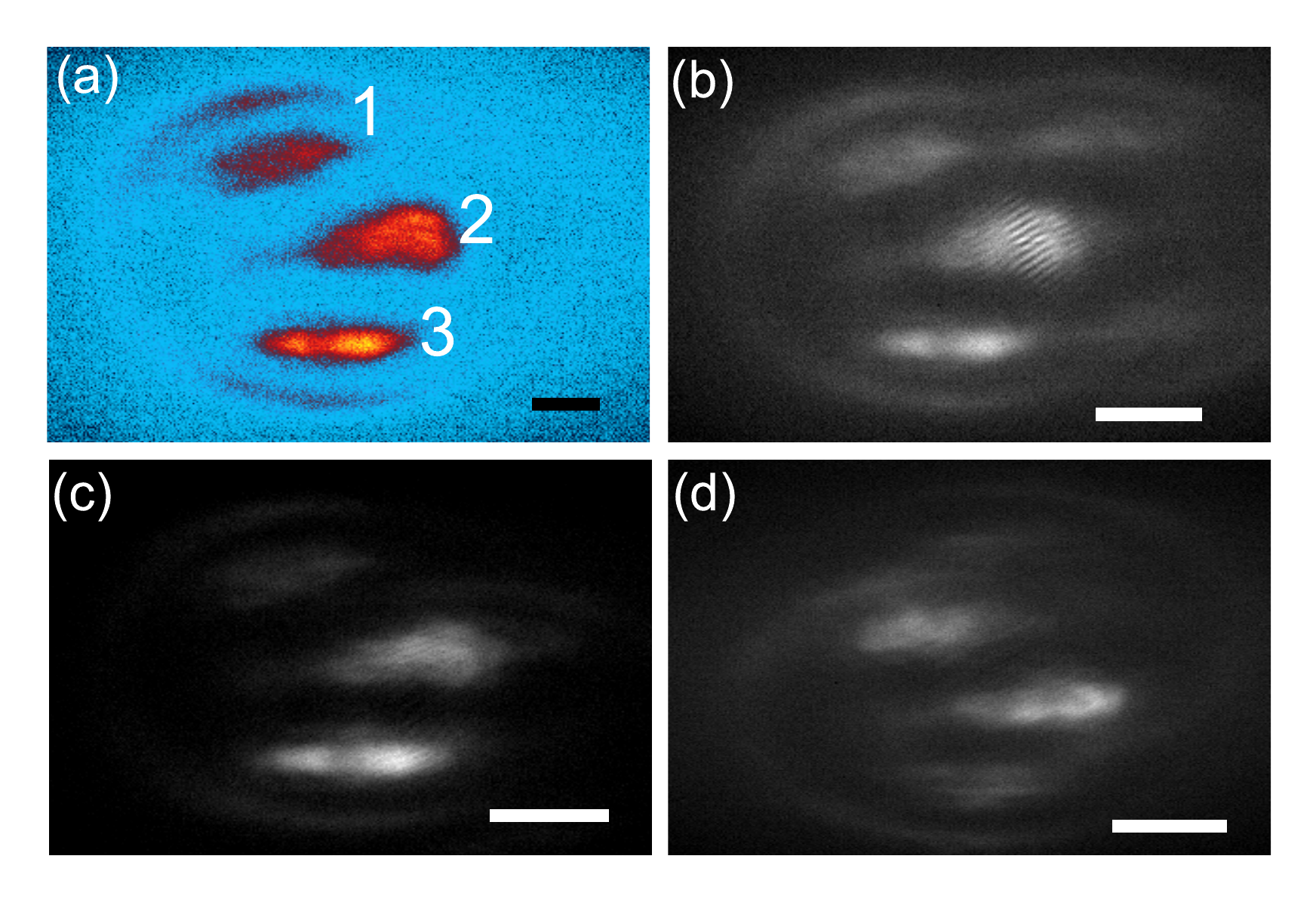}
\caption{\label{fig:supfig4} (color online). (a) Real-space image for a large pump showing 3 localized condensates denoted as 1, 2, 3. (b -- d) Interference pattern obtained by manually overlapping condensate 2 with itself (b), with 1 (c) and 3 (d). Scale bar is 15~$\mu$m.}
\end{figure}

In some cases, flat-top pumping leads to regions of high density separated by as much a ten microns. To probe whether these various regions are mutually coherent, we investigated a sample location, given in Fig.~\ref{fig:supfig4}(a), showing 3 seemingly isolated condensates denoted as 1, 2 and 3. By translating the mirror and retro-reflector images, we have performed interferometric experiments overlapping the central spot 2 with itself (b), with spot 1 (c) and spot 3 (d). Clear parallel fringes are observed in Fig.~\ref{fig:supfig4}(b) indicating that this condensate still shows strong phase correlations. However, no fringes can be resolved in Fig.~\ref{fig:supfig4}(b) and (c), which correspond to the interference between \emph{different} condensates. This evidences the presence of three distinct, but mutually incoherent condensates.

\end{document}